\documentclass[aps,reprint,twocolumn,showpacs,floatfix]{revtex4-1}
\usepackage{amsmath,bm,epsfig}
\usepackage{latexsym}
\usepackage{amssymb}
\usepackage{color}
\usepackage{morefloats}
\usepackage{subfigure}
\usepackage{graphicx}
\usepackage{amsmath}
\usepackage{bm}
\usepackage{epsf,epsfig,graphics}
\usepackage{float}
\usepackage{makeidx}
\usepackage{graphicx}
\usepackage{natbib}
\usepackage{epsfig}
\usepackage{rotate}
\usepackage{float}
\usepackage{amssymb}
\usepackage{amsbsy}
\usepackage{overpic}
\usepackage{marvosym}
\usepackage{textcomp}
\usepackage{amsmath}
\usepackage{longtable}
\usepackage{graphicx}
\usepackage{dcolumn}
\usepackage{bm}

\newcommand\Nu{\text{Nu}}
\newcommand\Sh{\text{Sh}}
\newcommand\Ra{\text{Ra}}
\newcommand\Rey{\text{Re}}
\newcommand\Pran{\text{Pr}}
\newcommand\Le{\text{Le}}
\newcommand\Sc{\text{Sc}}
\newcommand\dd{\partial }
\usepackage{color}

\newcommand{\blue}{\color{black} }

\newcommand{\bb}{\color{black} }

\begin{document}

\title{Buoyancy-driven melt rate of a vertical ice surface in seawater}
\author{Ho Yin Ng and Emily S.C. Ching}
\email{ching@phy.cuhk.edu.hk} \affiliation{Department of Physics,
The Chinese University of Hong Kong, Shatin, Hong Kong}
\date{\today}


\begin{abstract}

Reliable estimates of the melt rate of tidewater glaciers require accurate knowledge of salt and heat fluxes at the near-vertical ice-ocean interface and such knowledge is currently lacking. In this paper, we present a theory for salt and heat fluxes in the idealized system of turbulent double-diffusive convection in an infinite vertical channel and show how this theory can be extended to give theoretical estimate of the buoyancy-driven melt rate of a vertical ice surface in seawater. Our theoretical results are shown to be in good agreement with direct numerical simulation data and measurements from laboratory experiments.

\end{abstract}

\maketitle


\section{Introduction}
\label{intro}

Ice loss from ice sheets and glaciers is a dominant contributor to global sea level rise. The rate of ice loss from the Greenland and Antarctic ice sheets has increased rapidly in the past decades~\cite{PNAS1_2019,PNAS2_2019,nature1_2020,nature2_2018}. It has been demonstrated that much of the variability of ice loss from Greenland's tidewater glaciers is correlated with the warming of the ocean adjacent to the glaciers~\cite{PNAS2018} but estimating the rate of ice loss of tidewater glaciers and its dependence on ocean temperature remains a great challenge.  Accurate knowledge of the rates of salt and heat transfers at the vertical or near-vertical glacier faces in contact with the ocean is essential for a reliable estimate of the rate of ice loss from glaciers but this knowledge is currently lacking.

When formulating ice-ocean interactions, it is common to assume that
the temperature $T_i$ at an ice-ocean interface is equal to the freezing temperature $T_L({S}_i)$ of water with the salinity ${S}_i$ at the interface~\cite{HJ1999}:
\begin{equation}
T_i = T_L({S}_i) = -\lambda {S}_i
\label{Ti}
\end{equation}
where $\lambda$ is the slope of the liquidus curve. Heat flux from the warmer ambient ocean, at temperature $T_a$, provides energy to heat up and melt the ice. 
When the ice melts, freshwater at zero salinity is released and the salinity at the ice-ocean interface is maintained at ${S}_i$ by salt flux from the ambient seawater. Conservation of heat and salt at the interface implies~\cite{HJ1999}:
\begin{eqnarray}
 \rho_{s} V L +  k_{s} \frac{ \dd {T}}{\dd x} \bigg |_{x=0^-} &=& \rho_w c_w \kappa_T \Nu \frac{(T_a -{T}_i)}{h} \qquad  \label{heat} \\ 
\rho_{s} V {S}_i &= & \rho_w  \kappa_S \Sh \frac{(S_a -{S}_i)}{h} 
\label{salt}
\end{eqnarray}
Here, $x$ is the coordinate normal to the interface, with $x<0$ inside the glacier and $x>0$ in the ocean, $V$ is the ice melt rate, $\rho_{s}$, $L$ and $k_{s}$ are the density, specific latent heat of fusion and thermal conductivity of ice, $\rho_w$ and $c_w$ are the density and specific heat capacity of sea water, $\kappa_T$ and $\kappa_S$ are the thermal and salinity diffusivities of seawater, $S_a$ is the ambient salinity of the ocean, and $h$ is the underwater depth of the submerged glacier face. The dimensionless Nusselt  ($\Nu$) and  Sherwood numbers ($\Sh$) measure the heat and salt fluxes at the interface, and are defined by
\begin{equation}
\Nu =\frac{\dd \overline{T}/\dd x |_{x=0^+}}{\Delta T_c/l_c}, \qquad \Sh = \frac{\dd \overline{S}/\dd x |_{x=0^+}}{\Delta S_c/l_c}
\label{Nu}
\end{equation}
where an overbar denotes time average, $l_c$, $\Delta T_c$, and $\Delta S_c$ are characteristic length, salinity and temperature scales, which are taken to be $\Delta T _c=T_a-T_i$, $\Delta S_c=S_a-S_i$ and $l_c=h$, respectively for tidewater glaciers.
Approximating the glacier as a semi-infinite block receding at a constant rate $V$ gives the temperature distribution within the block~\cite{Wexler1960}:
\begin{equation}
T(x) = T_s + (T_i-T_s) e^{\rho_s c_s Vx/k_{s}}, \qquad x \le 0 , 
\label{Ts}
\end{equation}
where $T_s=T(-\infty)$ is the interior temperature of the glacier and $c_{s}$ is the specific heat capacity of ice. Using Eq.~(\ref{Ts}), Eq.~(\ref{heat}) can be rewritten as
\begin{eqnarray}
\rho_{s} V [L + c_{s} (T_i - T_s)] = \rho_w c_w \kappa_T  \Nu \frac{(T_a - T_i)}{h} 
\label{heat2}
\end{eqnarray}
Therefore, the melt rate $V$ of a tidewater glacier of submerged height $h$ and interior ice temperature $T_s$ in an ocean with ambient temperature $T_a$ and salinity $S_a$ is governed by Eqs.~(\ref{Ti}), (\ref{salt}) and (\ref{heat2}), and are thus determined by the salt and heat fluxes, measured by $\Sh$ and $\Nu$, at the ice-ocean interface. 
For the typically large values of $h$ of tidewater glaciers, convective flow driven solely by buoyancy due to opposing temperature and salinity differences between the interface and the ambient ocean is turbulent. There is not yet a theory for salt and heat fluxes in such turbulent double-diffusive convective flows. 
In geophysical studies, parameterizations of salt and heat fluxes in terms of turbulent transfer or exchange coefficients, a velocity scale and salinity and temperature differences are often employed~(see e.g.~\cite{Notz2003,Jenkins2010,Straneo2015}). It has been shown that an existing theory employing a common melt parameterization with typical coefficient values significantly underestimates the ambient melt rate of a tidewater glacier~\cite{Sutherland2019,Jackson2020}. 

Besides the melt rate $V$, the ratio of the dimensionless salt to heat fluxes:
\begin{equation}
R \equiv \frac{\Sh}{\Nu},
\label{R}
\end{equation}
which determines the salinity and temperature at the ice-ocean interface, 
is another quantity of interest~\cite{McPhee2008,Keitzl2016}.
The Sherwood and Nusselt numbers are inversely proportional to the thicknesses of the saline and thermal boundary layers $\delta_S$ and $\delta_T$, respectively. 
It has been proposed that the boundary layers grow diffusively with time, $\delta_S \sim \sqrt{\kappa_S t}$ and $\delta_T \sim \sqrt{\kappa_T t}$~\cite{Kerr1994}, and such an assumption leads to $R = \Le^{1/2}$, 
where
\begin{equation}
\Le = \frac{\kappa_T}{\kappa_S}
\label{Le}
\end{equation}
is the Lewis number. To shed light on the physical mechanisms that determine $R$, direct numerical simulations (DNS) of turbulent double diffusive convection in an infinite vertical channel between two walls at different temperatures and salt concentrations were performed at Schmidt number $\Sc =10$ and $100$~\cite{HVL2023}. Heavy computational cost makes it difficult to reach realistic values $\Sc$, which are greater than $2000$. The DNS study reveals a more complex dependence of $R$ on $\Le$: $R = \Le^{1/3}$ for $\Le <1$ and $\Sc=10$ and towards a steeper dependence closer to $\Le^{1/2}$ for larger $\Le$ and $\Sc=100$~\cite{HVL2023}. 

For the single-component turbulent convection in an infinite vertical channel between two walls at different temperatures, which will be referred to as turbulent vertical convection, we have  developed an eddy thermal diffusivity model that gives analytical results for the mean temperature profiles and an analytical estimate of the heat flux~\cite{JFM}. In this paper, we build upon this approach to derive analytical estimates for both the salt and heat fluxes for turbulent double-diffusive convection in an infinite vertical channel that is driven mainly by salt concentration difference, as in realistic polar oceans. Our theoretical results for $\Sh$ and $\Nu$ are in good agreement with the DNS results and explain the observed complex dependence of $R$ on $\Le$~\cite{HVL2023}. We further show that our theory can be extended to estimate the melt rate and temperature of an ice mass with a vertical ice face in saline water, and thus the buoyancy-driven melt rate of tidewater glaciers. Our theoretical estimates are in accord with laboratory measurements of the melting of a vertical ice wall in saline water~\cite{KM2015}. 

\section{Theory}
\subsection{Turbulent double-diffusive convection in an infinite vertical channel}
\label{theory}

\begin{figure}[h!]
    \centering
 \includegraphics[width=0.7\linewidth]{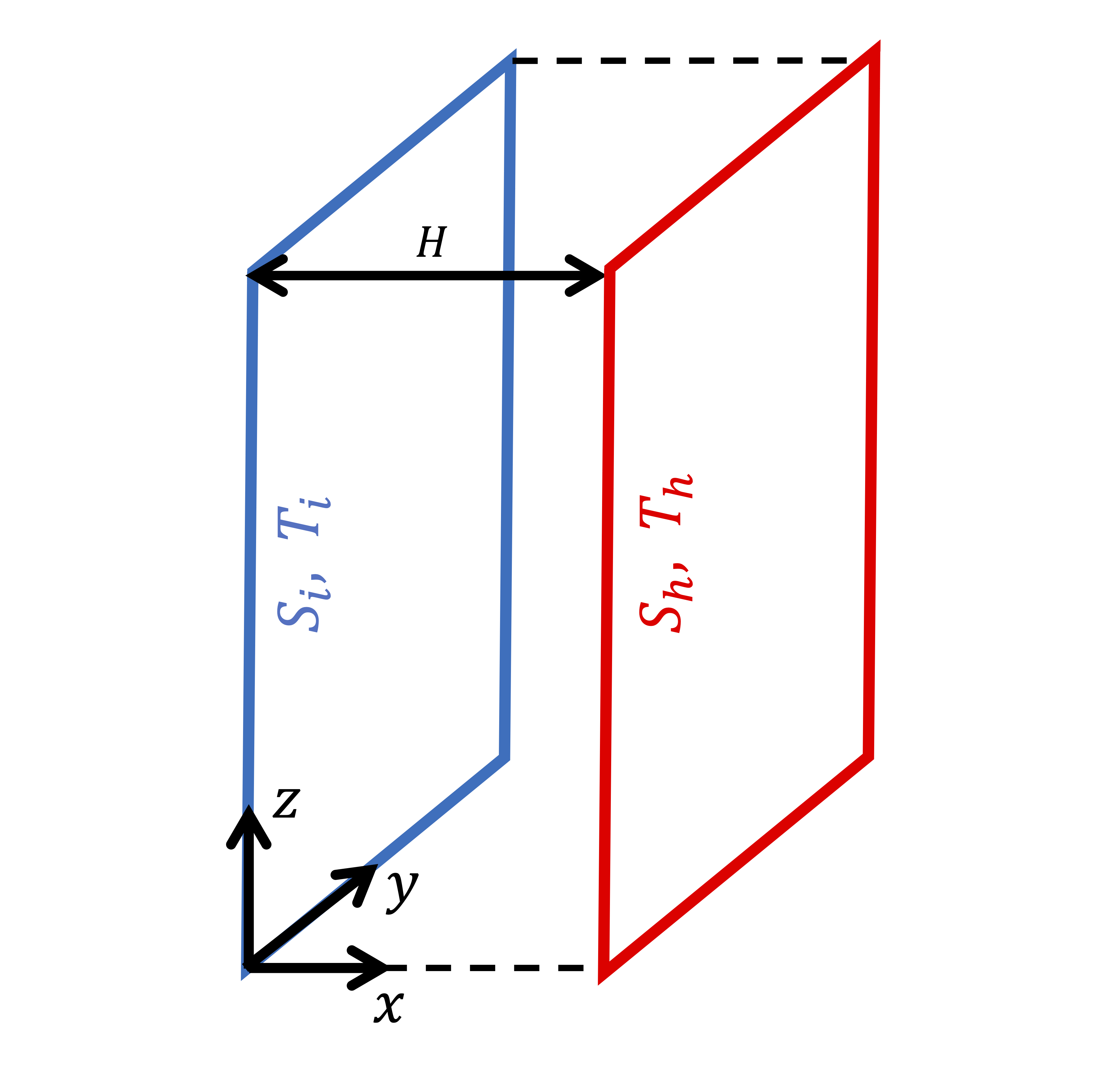}
    \caption{Schematic diagram of double-diffusive convection in an infinite vertical channel between two walls at different temperatures and salinities.}
    \label{Fig1}
\end{figure}

Consider turbulent double-diffusive convection between two infinite vertical walls, with the left wall at $x=0$ at temperature $T_i$ and salinity $S_i$, which mimics the ice-water interface, and the right wall at $x=H$ at a higher temperature $T_h$ and higher salinity $S_h$~(see Fig.~\ref{Fig1}). 
The governing equations of motion are:
\begin{eqnarray}
\nonumber    \frac{\partial \mathbf{u} }{\partial t}+{\mathbf{u}}\cdot\nabla {\mathbf{u}}&=&- \frac{1}{\rho_0} \mathbf{\nabla} p + \nu\nabla^2 {\mathbf{u}} \\
&& \hspace{0.2cm} +g [\alpha (T-T_0)- \beta(S-S_0)]\hat{z}  \label{MDDeqn} \\
 \frac{\partial S}{\partial t}+{\mathbf{u}}\cdot\nabla S&=&\kappa_S\nabla^2S \label{Ceqn} \\ 
    \frac{\partial T}{\partial t}+{\mathbf{u}}\cdot\nabla T&=&\kappa_T\nabla^2T \label{Teqn} \\  
    \nabla\cdot{\mathbf{u}}&=&0  \label{divfree}
\end{eqnarray}
and has been studied by direct numerical simulations~\cite{HVL2023}.
Here, $\mathbf{u}=(u,v,w)$ is the velocity field, $p$ the pressure field, $T$ and $S$ are the temperature and salinity,  
 $\rho_0$ is the density of water at temperature $T_0$ and salinity $S_0$, which are the average temperature and salinity of the two walls, $\alpha$, $\beta$ and $\nu$ are the volume expansion coefficient, haline contraction coefficient and kinematic viscosity of the saline water, and $\hat{z}$ is a unit vector along the $z$-direction~(see Fig.~\ref{Fig1}). We have adopted the Oberbeck-Boussinesq approximation in which the equation of state is $\rho = \rho_0[1-\alpha(T-T_0)+\beta(S-S_0)]$ and variations of fluid properties with temperature and salinity are negligible except for generating the buoyancy term in the momentum equation Eq.~(\ref{MDDeqn}). At the two vertical walls, the velocity field satisfies no slip boundary condition.
There are four independent dimensionless parameters, which are chosen as the density ratio $R_\rho$,  
salinity Rayleigh number $\Ra_S$, Schmidt number $\Sc$ and Prandtl number $\Pran$, and their definitions are:
\begin{eqnarray}
R_\rho = \frac{\alpha\Delta T_c}{\beta \Delta S_c}; \  \Ra_S= \frac{\beta g \Delta S_c l_c^3}{\nu \kappa_S}; \  
\Sc =\frac{ \nu}{\kappa_S}; \  \Pran = \frac{\nu}{\kappa_T} \ \
\label{def}
\end{eqnarray} 
where $\Delta T_c=T_h-T_i$, $\Delta S_c=S_h-S_i$ and $l_c=H$.
Taking the time average of the governing equations leads to
 \begin{eqnarray}
    \frac{d}{dx} \overline{u'w'} &=&\nu \frac{d^2\overline{w}}{dx^2}  + g[\alpha (\overline{T}-T_0) - \beta (\overline{S}-S_0)] \qquad   \label{MBEDD}\\
 \frac{d}{dx} \overline{u'S'}&=&\kappa_S \frac{d^2 \overline{S}}{dx^2}  \label{MCEDD}  \\
      \frac{d}{dx} \overline{u'T'}&=&\kappa_T \frac{d^2 \overline{T}}{dx^2}
    \label{MTEDD}
\end{eqnarray}
where the time-averaged flow quantities depend on $x$ only as the two vertical walls are infinite and the primed quantities are the fluctuations.
Because of the symmetry of the setup, $\overline{w}(x)$, 
$\overline{S}(x)-S_0$ and $\overline{T}(x)-T_0$  are antisymmetric about $x=H/2$ and thus vanish at $x=H/2$. 

The quantities of interest are the dimensionless salt and heat fluxes, $\Sh$ and $\Nu$ as defined in Eq.~(\ref{Nu}). We first show that $\Sh$ and $\Nu$ can each be expressed solely in terms of their respective space-dependent turbulent or eddy diffusivities, the eddy saline diffusivity $K_S(x)$ and eddy thermal diffusivity $K_T(x)$, defined by
\begin{eqnarray}
\overline{u' S'} \equiv - K_S(x)  \frac{d \overline{S}}{dx}, \qquad \overline{u' T'} \equiv - K_T(x)  \frac{d \overline{T}}{dx}  
\label{KSKT} 
\end{eqnarray}
Substitute~Eq.~(\ref{KSKT}) into~Eqs.~(\ref{MCEDD}) and (\ref{MTEDD}) and integrate the resulting equations, we obtain
 \begin{eqnarray}
 \Sh &=& \frac{1}{2} \left[ \int_0^{1/2} \frac{dy}{1+\Sc \ K_S(yH)/\nu} \right]^{-1} \label{NuS} \\
 \Nu &=& \frac{1}{2} \left[ \int_0^{1/2} \frac{dy}{1+\Pran \ K_T(yH)/\nu} \right]^{-1} \label{NuT} 
 \end{eqnarray}
 where we have used the boundary conditions for $\overline{S}(0)=S_i$, $\overline{S}(H/2)=S_0$, $\overline{T}(0)=T_i$ and $\overline{T}(H/2)=T_0$.
Next we show that $K_T(x) \approx K_S(x)$ at high $\Ra_S$. This approximation has been used in the literature~(see e.g., \cite{Gade1979}) and was supported by numerical simulations~\cite{Smyth2005,MP2022,HVL2023,Couston2024} and laboratory experiments~\cite{JR2003,MR2006} but,  to the best of our knowledge, a theoretical justification has not been presented.
Using~Eqs.~(\ref{Ceqn}) and (\ref{Teqn}), we obtain
 \begin{eqnarray}
\frac{D}{Dt} S' + \frac{\partial}{\partial x_i} \left(u'_i S' - \kappa_S \frac{\partial S'}{\partial x_i} - \overline{u'_i S'} \right) 
&=& - u' \frac{d \overline{S}}{dx}  \label{psi'} \qquad  \\
\frac{D}{Dt} T' +\frac{\partial}{\partial x_i} \left(u'_i T' - \kappa_T \frac{\partial \theta'}{\partial x_i} - \overline{u'_i T'} \right) 
&=& - u' \frac{d \overline{T}}{dx} \label{theta'} \qquad \end{eqnarray}
where $D/Dt \equiv \partial/\partial t + \overline{u}_i \partial/\partial x_i$ and we have used Einstein's summation convention and the notations: $\vec{r}=(x,y,z)=(x_1,x_2,x_3)$, $(\overline{u}, \overline{v}, \overline{w})=(\overline{u}_1, \overline{u}_2,\overline{u}_3)$ and $(u',v',w')=(u'_1,u'_2,u'_3)$. 
At high $\Ra_S$, turbulent fluxes are expected to dominate molecular diffusive fluxes, i.e.,
\begin{eqnarray}   
|u'_i S'| \gg  \kappa_S \left|\frac{\partial S'}{\partial x_i}\right|  {\rm and} \
|u'_i T'| \gg \kappa_T \left| \frac{\partial T'}{\partial x_i}\right|
\end{eqnarray}
As a result, Eqs.~(\ref{psi'}) and (\ref{theta'}) can be approximated as
\begin{eqnarray}
\frac{D}{Dt} S' + \frac{\partial}{\partial x_i} \left(u'_iS'  - \overline{u'_i S' }\right)  &\approx& - u' \frac{d \overline{S}}{dx} \label{psi'approx} \\
\frac{D}{Dt} T' +\frac{\partial}{\partial x_i} \left(u'_i T' - \overline{u'_i T'}\right) 
&\approx& - u' \frac{d \overline{T}}{dx} \label{theta'approx}
\end{eqnarray}
which show that the scalar fluctuation, $S'$ or $T'$, is determined by the velocity fluctuations $u'_i$ and the mean scalar gradient, $d\overline{S}/dx$ or $d\overline{T}/dx$. 
Following the work of Hamba~\cite{Hamba1995}, we introduce the Green's function $G(\vec{r},t;\vec{r'},t')$, which satisfies
\begin{widetext}
\begin{eqnarray}
 \frac{D}{Dt} G(\vec{r},t;\vec{r'},t') + \frac{\partial}{\partial x_j}[u'_j(\vec{r},t)G(\vec{r},t;\vec{r'},t') - \overline{u'_j(\vec{r},t)G(\vec{r},t;\vec{r'},t')}] 
= u'(\vec{r}',t') \delta(x-x')\delta(y-y')\delta(z-z')\delta(t-t') \ \
\label{Green}
\end{eqnarray}
Then
\begin{eqnarray}
S'(\vec{r},t) \approx - \int_0^{H/2} \int_{-\infty}^{\infty} \int_{-\infty}^{\infty} \int_{-\infty}^t G(\vec{r},t;\vec{r'},t') \frac{d \overline{S}}{dx'}  dt' dy' dz' dx' 
\approx - {g}(\vec{r},t) \frac{d \overline{S}}{dx} \label{psi2} \\
T'(\vec{r},t) \approx - \int_0^{H/2} \int_{-\infty}^{\infty} \int_{-\infty}^{\infty}  \int_{-\infty}^t G(\vec{r},t;\vec{r'},t') \frac{d \overline{T}}{dx'}  dt' dy' dz' dx' \approx 
 - {g}(\vec{r},t) \frac{d \overline{T}}{dx} \label{theta2}
\end{eqnarray}
\end{widetext}
where
\begin{eqnarray}
{g}(\vec{r},t) \equiv \int_{|\vec{r'}-\vec{r}| \le \delta} \int_{-\infty}^t G(\vec{r},t;\vec{r'},t')  dt' d\vec{r'}  \ \ 
\end{eqnarray}
When evaluating the integrals in Eqs.~(\ref{psi2}) and (\ref{theta2}), we have assumed that $G(\vec{r},t;\vec{r'},t') \approx 0$ for $|\vec{r}-\vec{r'}|$ exceeding the turbulent length scale $\delta$ and  
$d\overline{S}/dx'\approx d\overline{S}/dx$  or $d\overline{T}/dx'\approx d\overline{T}/dx$ for $|x'-x| \le\delta$.
Equations~(\ref{psi2}) and (\ref{theta2}) imply
\begin{eqnarray}
 \overline{u' S'} \approx -  \overline{u'g} \  \frac{d \overline{S}}{dx}, \qquad
\overline{u' T'}\approx -  \overline{u'g} \ \frac{d \overline{T}}{dx} \qquad  \label{approx} 
\end{eqnarray}
which yields 
\begin{eqnarray}
 K_T(x) \approx K_S(x) &\approx&  \overline{u' {g}}
\label{KTKS}
\end{eqnarray}  
Thus, Eq.~(\ref{NuT}) becomes
\begin{eqnarray}
 \Nu \approx \frac{1}{2} \left[ \int_0^{1/2} \frac{dy}{1+\Pran \ K_S(yH)/\nu} \right]^{-1} \label{NuT2} 
 \end{eqnarray}
 As a result, both $\Sh$ and $\Nu$ can be determined if the space-dependent eddy saline diffusivity $K_S(x)$ is known.
 
In polar oceans, $R_\rho \sim (\alpha/\beta) (T_a-T_i)/(S_a-S_i) < 10^{-2}$ such that the convective flow is essentially driven by salinity difference.
In such convective flow, which we will consider only in this paper, the salinity field plays the role of an active scalar like the temperature field in turbulent vertical convection, with $\Ra_S$ and $\Sc$ corresponding to the thermal Rayleigh number $\Ra=\alpha g (T_h-T_i) H^3/(\nu \kappa_T)$ and $\Pran$. Thus, we adopt the three-layer eddy diffusivity model that we recently proposed for turbulent vertical convection~\cite{JFM} for $K_S(x)$:
 \begin{eqnarray}
\frac{K_S(yH)}{\nu} = \begin{cases} A y^3, & 0 \le y \le y_1  \\
c_1 + c_2 y, &  y_1 < y < y_2  \\
C_m[1- (1-2y)^2], & y_2 \le y \le 1/2 
\end{cases}
\label{model} , \ \ 
\end{eqnarray}
Here, $y=x/H$, $y_1=2 l_i/H$ with $l_i=H/({\rm Sc}A)^{1/3}$, $y_2=0.3$ and $A$ and $C_m$ are two parameters which determine the characteristic velocities for salinity transfer in the inner wall and outer regions~\cite{JFM}. The cubic dependence near the wall follows from the requirements that $K_S(x)$ and its first- and second-order derivatives vanish at $x=0$, which are imposed by the boundary conditions, and the
quadratic functional form near $x=H/2$ is based on the symmetry of $K_S(x)$ about the centerline $x=H/2$~\cite{JFM}. 
The constants $c_1$ and $c_2$ are related to $A$ and $C_m$ by the imposed continuity of $K_S(yH)/\nu$ at $y=y_1$ and $y_2$.
The values of $A$ and $C_m$ for moderate values of $\Ra_S$ and $\Sc$ were extracted directly from the DNS data of turbulent vertical convection~\cite{HNVL2022}, 
and their dependence on $\Ra_S$ and $\Sc$ in the high-$\Ra_S$ limit are estimated~\cite{JFM} as
\begin{eqnarray}
A &=&  a_1 \Sc^{-34/21} \Ra_S \label{A} \\
C_m&=& a_2 \Sc^{-19/27} \Ra_S^{4/9}  \label{Cm}
\end{eqnarray}
with $a_1=0.0132$, $a_2=0.0669$.
Applying~Eq.~(\ref{model}) to~Eqs.~(\ref{NuS}) and (\ref{NuT2}) yields~\cite{JFM}:
\begin{eqnarray}
\Sh &=& \frac{1}{2 [I_1(\Sc)+ I_2(\Sc) + I_3(\Sc)]} \label{NuSeqn} \\
\Nu &\approx& \frac{1}{2[I_1(\Pran)+ I_2(\Pran) + I_3(\Pran)]}  
\label{NuTeqn}
\end{eqnarray}
where 
\begin{widetext} 
\begin{eqnarray} I_1(x) &=& \int_0^{y_1}  \frac{dy}{1+x \ K_S(yH)/\nu} = 
 \frac{1}{3x^{\frac{1}{3}} A^{\frac{1}{3}}} \left\{ \frac{1}{2} \log\left[\frac{(1+2x^{1/3}/\Sc^{1/3})^3}{1+ 8x/\Sc}\right] 
 + \sqrt{3}\arctan \left[\frac{4(x/\Sc)^{1/3} -1}{\sqrt{3}} \right]+ \frac{\sqrt{3} \pi}{6} \right\} \label{I1},  \qquad \\
 I_2(x) &=&  \int_{y_1}^{y_2} \frac{dy}{1+x \ K_S(yH)/\nu}  =
\frac{1}{x c_2} \log \left[ \frac{1+8x/\Sc}{1+21/25 x C_m} \right] \label{I2}, \\
 I_3 (x) &=& \int_{y_2}^{1/2} \frac{dy}{1+x \ K_S(yH)/\nu}  
= \frac{1}{4\sqrt{x C_m(1+x C_m )}}\log \left| \frac{5\sqrt{1+x C_m} +2\sqrt{x C_m}}{5\sqrt{1+x C_m} - 2\sqrt{ x C_m}} \right| \qquad
\label{I3}
\end{eqnarray} 
\end{widetext}
Hence, we obtain analytical estimates of $\Sh$, $\Nu$ and thus $R$ for different values of $\Ra_S$, $\Sc$, and $\Pran$.

\subsection{Turbulent double-diffusive convection adjacent to a vertical ice face in saline water}

For an ice mass with a vertical ice face at $x=0$ in saline water at temperature $T_a$ and salinity $S_a$, the system is governed by the same equations~[Eqs.~(\ref{MDDeqn})-(\ref{divfree})] for $x \ge 0$ with $T_0 = T_a$ and $\rho_0 = \rho_w$, and the boundary conditions are ${\bf u}=0$ at $x=0$ and $T \to T_a$ and $S \to S_a$ as $x \to \infty$. Here, the fluid flow is studied in a coordinate system that moves together with the vertical ice face at the melt rate $V$, which is taken to be approximately constant. The characteristic scales are now $\Delta T_c=T_a-T_i$, $\Delta S_c=S_a-S_i$ and $l_c = h$, where $h$ is the underwater depth of the submerged ice surface, and $T_i$ and $S_i$ are the mean temperature and salinity at the ice surface to be solved using Eqs.~(\ref{Ti}), (\ref{salt}) and Eq.~(\ref{heat2})~(see Sec.~\ref{intro}). For this system, the mean flow quantities are functions of $(x, y, z)$ but in the flow region adjacent to the ice face, variations along the ice-water interface are much smaller than variations normal to the interface, namely $\partial/\partial y \ll \partial/\partial x$ and $\partial/\partial z \ll \partial/\partial x$, as in usual boundary layers. Moreover, as $\Sc$ is of the order of $10^3$, the saline boundary layer is thin and the mean velocity components can be neglected inside the boundary layer. Therefore, 
\begin{eqnarray}
\frac{\partial}{\partial x} \overline{u'S'} \approx \kappa_S \frac{ \partial^2 \overline{S}}{\partial x^2}, \qquad 
\frac{\partial}{\partial x} \overline{u'T'} \approx \kappa_T \frac{\partial^2 \overline{T}}{\partial x^2} \label{appeqn}
\end{eqnarray}
for $0 \le x \le d$ near the ice-water surface. We can thus extend the theory for an infinite vertical channel to obtain
\begin{eqnarray}
 \Sh &\approx& \mu \left[I_1(\Sc)+I_2(\Sc)+I_3(\Sc)\right]^{-1} \label{Sh_single} \qquad \\
 \Nu &\approx&  \mu   \left[I_1(\Pran)+I_2(\Pran)+I_3(\Pran)\right]^{-1} \label{Nu_single}  \qquad 
 \end{eqnarray}
where 
\begin{equation}
\mu = \left( \frac{l_c}{2d}  \right) \left[ \frac{\overline{S}(d)-S_i}{\Delta S_c} \right]
\label{mu}
\end{equation} and we have made the approximation
\begin{eqnarray}
\frac{\overline{T}(d)-T_i}{\Delta T_c}  \approx  \frac{\overline{S}(d)-S_i}{\Delta S_c} 
\label{app}
\end{eqnarray}
Recall that for an infinite vertical channel of separation $H$, Eq.~(\ref{appeqn}) with $\partial/\partial x$ replaced by $d/dx$, are exact and hold for the whole channel giving $2d=H=l_c$ and the symmetry about $x=H/2$ leads to $\overline{S}(d)-S_i = \Delta S_c/2$ and $\overline{T}(d)-T_i=\Delta T_c/2$ and thus the approximation Eq.~(\ref{app}) becomes exact and $\mu=1/2$. As a result, Eqs.~(\ref{Sh_single}) and (\ref{Nu_single}) recover Eqs.~(\ref{NuSeqn}) and (\ref{NuTeqn}). For the complex case of an ice mass submerged in saline or sea water, there is an additional parameter $\mu$ when calculating $\Sh$ and $\Nu$. By assuming $\mu$ to be a constant and matching the dependence of $\Sh$ on $\Ra_S$ in the high-$\Ra_S$-limit with  experimental results for heat flux from a heated vertical plate in air~($\Nu=\gamma^{\rm air} \Ra^{1/3}$ with $\gamma^{\rm air} \sim 0.1-0.12$)~\cite{TN1988,Holman2010}, we obtain $\mu = 0.5 \pm 0.07$. Therefore, we estimate $\mu \approx 1/2$.
Dividing~Eq.~(\ref{heat2}) by~Eq.~(\ref{salt}) yields 
\begin{equation}
\frac{R}{c_w \Le} [L + c_{s} (T_i- T_s)](S_a - S_i)  =  S_i (T_a -T_i)
\label{eqSi}
\end{equation}
Substituting Eq.~(\ref{Ti}) into Eq.~(\ref{eqSi}), we obtain a quadratic equation for $S_i$ in terms of $R$, $T_s$, $T_a$ and $S_a$. It is important to note that since $A$ and $C_m$ depend on $\Ra_S$ and thus on $S_i$, $\Sh$, $\Nu$ and thus $R$ are functions of $S_i$. Hence, Eq.~(\ref{eqSi}) is an implicit equation for $S_i$ and has to be solved self-consistently. With $S_i$ and Eq.~(\ref{salt}), the melt rate of the ice face is obtained: 
\begin{equation}
V = \frac{\rho_w}{\rho_{s}} \frac{\kappa_S}{h}  \frac{(S_a-S_i)}{S_i} \Sh
\label{V}
\end{equation}
Hence, we have theoretical estimates of $T_i$ or $S_i$ and $V$ for given ambient salinity and  temperature $S_a$ and $T_a$, underwater depth $h$ and interior temperature $T_s$ of the ice mass.

\section{Results and Discussion}

\subsection{Salt and heat fluxes in turbulent double-diffusive convection}

A DNS study of turbulent double-diffusive convection in an infinite vertical channel was carried out by Howland et al.~\cite{HVL2023} for (i) $\Ra_S=10^7$, $\Sc=10$ and $\Pran=1, 2, 5, 20, 50, 100$ and (ii) $\Ra_S=10^8$, $\Sc=100$ and $\Pran=1, 2, 5, 10$ at a fixed density ratio of $R_\rho=0.02$. In Table~\ref{tab1}, we compare our theoretical results of $\Sh$ and $\Nu$, obtained using the values of $A$ and $C_m$ from Table~1 of~\cite{JFM}, with the DNS data. The model Eq.~(\ref{model}) for the eddy salinity diffusivity $K_S(x)$, which is equivalently the eddy active scalar diffusivity, gives good estimates of $\Sh$ with less than or around 5\% errors. While the estimates for $\Nu$ are generally less accurate, which is likely due to the additional assumption of $K_T(x) \approx K_S(x)$, our analytical results of $\Sh$ and $\Nu$ enable us to understand the complex dependence of $R$ on $\Le$.

\begin{widetext}

\begin{table*}[h]
\centering
\begin{tabular}{|c|c|c|c|c|c|c|c|c|c|}
\hline
$\Ra_S$ & $\Sc$  & $A$ & $C_m$ & $\Pran$ & $\Sh$ & $\Nu$ & $\Sh$ (DNS)  & $\Nu$ (DNS)  \\\hline
$10^7$ & 10 & 3955.05 & 17.98 & 1 & 13.18 (5.10\%) & 5.07 (9.27\%) & 12.54& 4.64   \\\hline
$10^7$ & 10 & 3955.05 & 17.98  & 2 &  13.18 (4.69\%) & 6.91 (7.30\%) &12.59 & 6.44   \\\hline
$10^7$ & 10 & 3955.05 & 17.98  & 5 & 13.18 (4.27\%) & 10.09 (5.54\%) & 12.64 & 9.56  \\\hline
$10^7$ & 10 & 3955.05 & 17.98  & 20 & 13.18 (3.70\%) & 17.01 (3.40\%) & 12.71 & 16.45   \\\hline
$10^7$ & 10 & 3955.05 & 17.98  & 50 & 13.18 (3.29\%) & 23.53 (4.35\%) & 12.76 & 22.55   \\\hline
$10^7$ & 10 & 3955.05 & 17.98  & 100 & 13.18 (3.53\%) & 29.90 (6.41\%) & 12.73 & 28.10   \\\hline 
$10^8$ & 100 & 784.51 & 7.00 & 1 & 18.12 (2.58\%) & 3.27 (11.99\%) & 18.60 & 2.92   \\\hline
$10^8$ & 100 & 784.51 & 7.00 & 2 & 18.12 (2.58\%) & 4.49 (13.10\%) & 18.60 & 3.97   \\\hline
$10^8$ & 100 & 784.51 & 7.00 & 5 & 18.12 (3.82\%) & 6.61 (8.01\%)  & 18.84 & 6.12  \\\hline
$10^8$ & 100 & 784.51 & 7.00 & 10 & 18.12 (4.68\%) & 8.56 (2.64\%)  & 19.01 & 8.34 \\\hline
\end{tabular}
\caption{Theoretical results of $\Sh$ and $\Nu$.  The numbers in parentheses are the absolute percentage errors with respect to the DNS data, where $\Sh$ and $\Nu$ are denoted as $\Nu_S$ and $\Nu_T$~\cite{HVL2023}.}
\label{tab1}
\end{table*}

\end{widetext}

\begin{figure}[h!]
    \centering
    \includegraphics[width=1.05\linewidth]{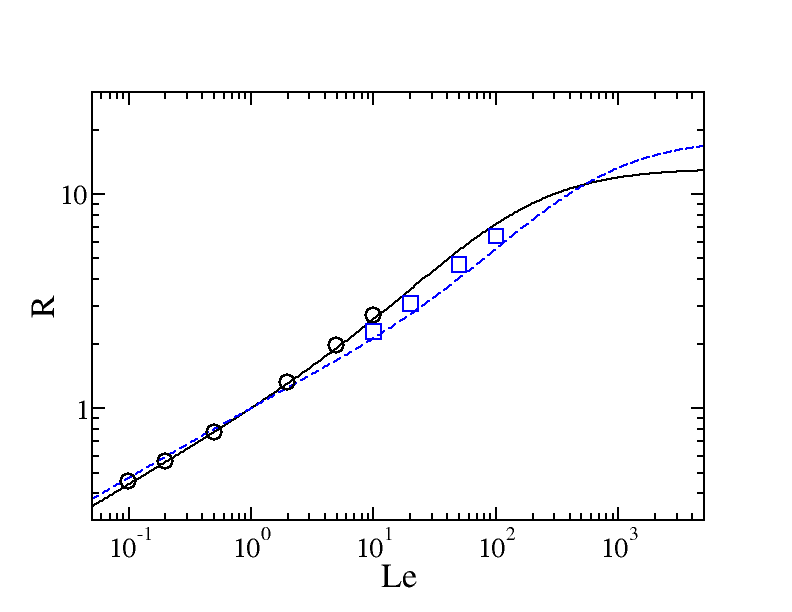}
    \caption{Flux ratio $R$ as a function of $\Le$ for $\Ra_S=10^7$ and $\Sc=10$~(solid line) and $\Ra_S=10^8$ and $\Sc=100$~(dashed line). The DNS data for $\Ra_S=10^7$~(circles) and $\Ra_S=10^8$~(squares)~\cite{HVL2023} are included for comparison.}
    \label{Fig2}
\end{figure}

Using Eqs.~(\ref{I1})-(\ref{I3}), it can be shown that at fixed $\Ra_S$ and $\Sc$, and thus fixed $A$, $C_m$ and $\Sh$, 
 $I_1(\Pran) \sim \Pran^{-1/3}$, $I_2(\Pran)\sim \Pran^{-1}$ for $\Le \ll 1$, $I_3(\Pran) \sim \Pran^{-1}$ for $\Pran C_m \gg 1$.  Equation~(\ref{NuT2})  implies that $\Nu \to 1$  as  $\Pran \to 0$. Hence, at fixed $\Ra_S$ and $\Sc$, 
\begin{eqnarray} 
R &\sim& \Le^{1/3} , \qquad \qquad \qquad \Le \ll 1 \label{case1} \\
R & \approx& a ( \Le^{1/3} + b \  \Le),  \ \ \ \mbox{intermediate $\Le$} \label{case2}\\
R &\to& c, \qquad \qquad \qquad \qquad  \  \Le \to \infty \label{case3}
\end{eqnarray}
where $a$, $b$ and $c$ are constants that depend on $\Ra_S$ and $\Sc$. Thus, $R$ depends on $\Ra_S$ and $\Sc$ in addition to $\Le$ and its dependence on $\Le$ at fixed $\Ra_S$ and $\Sc$ varies with $\Le$. In Fig.~\ref{Fig2}, we show the dependence of the theoretical results of $R$ on $\Le$. The theoretical results are consistent with the DNS data. The result of $R \sim \Le^{1/3}$ for small $\Le$, Eq.~(\ref{case1}), follows from the cubic dependence of $K_S(x)$ and was indeed observed~\cite{HVL2023}.  When the dependence of $R$ on $\Le$  for intermediate values of $\Le$ is fitted by a power law, Eq.~(\ref{case2}) implies that the effective exponent would have a  value between 1/3 and 1, and this explains the observed increase of the effective exponent towards $1/2$ around $\Le = 100$ at $\Ra_S=10^8$ and $\Sc=100$~\cite{HVL2023}.

\subsection{Melt rate of a vertical ice wall in saline water}

\begin{figure}[h]
    \centering
  \includegraphics[width=1.05\linewidth]{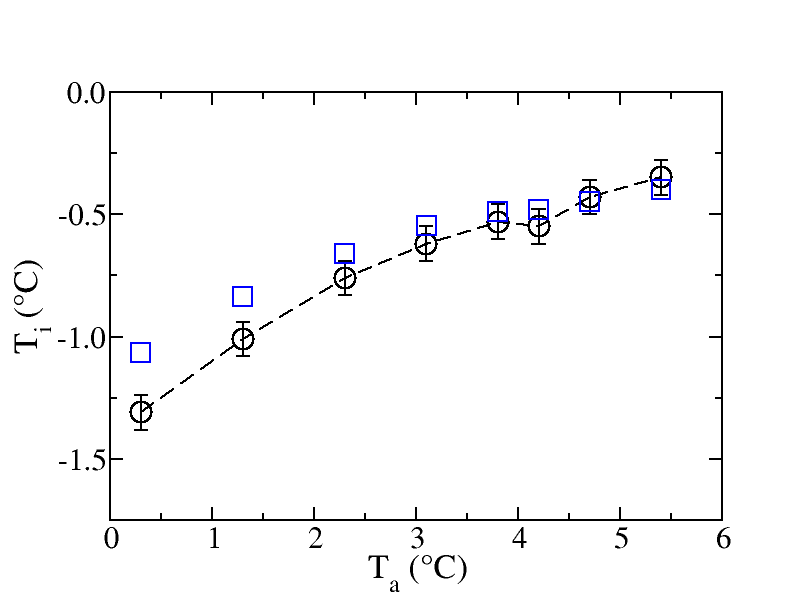}
    \caption{Comparison of the theoretical results of the temperature $T_i$ at the ice interface, averaged over the four different heights,~(squares) with the experimental measurements $T_w$~(circles) with an error bar of $0.1^\circ$C~\cite{KM2015}. }
    \label{Fig3}
\end{figure}

\begin{figure}[h]
    \centering
   \includegraphics[width=1.05\linewidth]{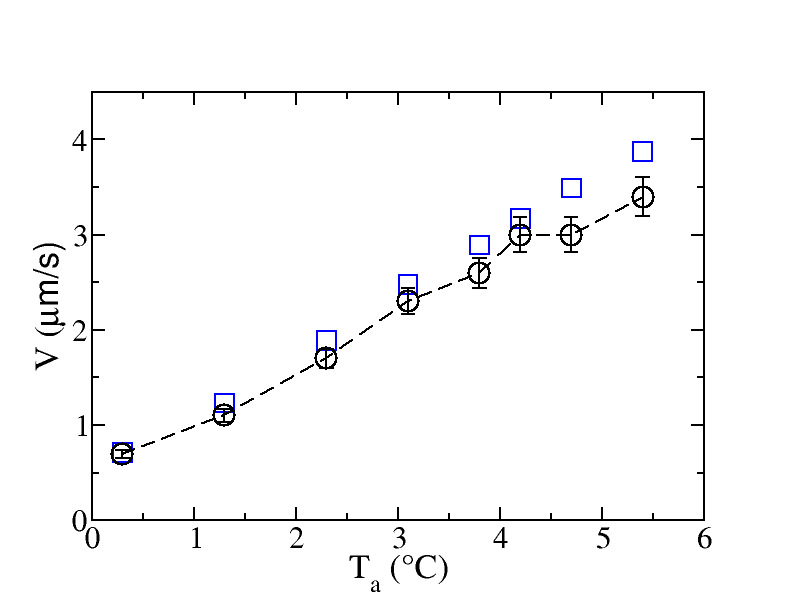}
    \caption{Comparison of the theoretical results of the melt rate $V$, averaged over the four different heights,~(squares) with the experimental measurement $V_m$~(circles) with an error bar of the standard deviation of 6\%~\cite{KM2015}. }
    \label{Fig4}
\end{figure}

The ablation or melt rate of a vertical ice wall in contact with saline water has been studied in laboratory experiments~\cite{JM1981,KM2015}. We will focus on the experiments by Kerr and McConnochie~\cite{KM2015} in which a fresh-water ice wall was at one end of a tank filled with homogeneous aqueous solutions of sodium chloride at temperature $T_a$ and salinity $S_a$, and the ablation rate of the ice-water interface was measured.  Shadowgraph showed that the flow became turbulent when $h > 20$~cm. Measured temperatures $T_w$ at the ice-water interface at four different heights, $h=0.32$~m, 0.42~m, 0.56~m and 0.70~m, were found to be constant to within 0.1$^\circ$C.  Using the measured melt rate $V_m$ and the measured rate of temperature rise $d{T}/dt$ in the ice near the interface, Kerr and McConnochie estimated the equivalent temperature difference $T_\delta=T_i-T_s$ as
\begin{equation}
  T_{\delta}\approx \frac{dT}{dt} \frac{k_s}{\rho_s c_s V_m^2}
\label{TiTs}
\end{equation}
In terms of $T_\delta$, Eq.~(\ref{eqSi}) can be rewritten as
\begin{equation}
\frac{R}{c_w \Le} (L + c_{s}T_{\delta})(S_a - S_i)  =  S_i (T_a + \lambda S_i)
\label{eqSi2}
\end{equation}
using Eq.~(\ref{Ti}). Using our theory for $\Sh$ and $\Nu$ and thus $R$, Eq.~(\ref{eqSi2}) can again be solved consistently to obtain $S_i$ and together with Eq.~(\ref{V}), $V$ is obtained. 
We show our theoretical results in Table~\ref{tab2} together with the parameters and measurements of the experiments, extracted from Table 3 of \cite{KM2015}. Our theoretical results of $T_i$ and $V$ depend on $h$ but the differences in the values for the four heights are well within the measurement errors. In Figs.~\ref{Fig3} and \ref{Fig4}, we compare our theoretical results of $T_i$ and $V$, averaged over the four different heights, with the experimental measurements. Good agreement can be seen.

\begin{widetext}

\begin{table*}[h!]
\centering
\begin{tabular}{|c|c|c|c|c|c|c||c|c|c|c|c|c|c|c|}
\hline
 & $T_a$ &$S_a$  &  $\rho_w$ & $T_\delta$ & $T_w$ & $V_m$ & \multicolumn{2}{c|} {$h$=0.32~m} & \multicolumn{2}{c|} {$h$=0.42~m} & \multicolumn{2}{c|} {$h$=0.56~m} 
& \multicolumn{2}{c|} {$h$=0.70~m} \\ \cline{8-9} \cline{8-15} 
 &  ($^\circ$C) &  (g/kg)  &  (kg/m$^3$) & ($^\circ$C) &  ($^\circ$C) & ($\mu$m/s)& $T_i$ ($^\circ$C) & $V$ ($\mu$m/s)& $T_i$ ($^\circ$C) & $V$ ($\mu$m/s) & $T_i$ ($^\circ$C) & $V$ ($\mu$m/s) & $T_i$ ($^\circ$C) & $V$ ($\mu$m/s) \\ \hline 
A & 0.3 & 34.4 &1026.1 &24 &-1.31 &0.7 & -1.073  & 0.692 &  -1.067 &   0.703  & -1.060 &  0.714  &-1.055 &   0.724 \\ \hline
B &1.3 &34.9 &1026.4 &20  &-1.01 & 1.1  & -0.847  & 1.202  & -0.840 &  1.222 & -0.833  & 1.244 & -0.827 &  1.261  \\ \hline
C &2.3 &35.0 &1026.4 &11 &-0.76 &1.7 & -0.669  & 1.836 & -0.663  & 1.867  &-0.655   &1.902&  -0.650  & 1.930  \\ \hline
D &3.1 &34.7 &1026.1 &4 &-0.62 &2.3 & -0.557   &2.408 & -0.550 &  2.450 & -0.544  & 2.497 & -0.538  & 2.535  \\ \hline
E &3.8 &34.6 &1026.0 &7 &-0.53 &2.6 & -0.498  & 2.816  &-0.493  & 2.866  &-0.486  & 2.921 & -0.481 &  2.921  \\ \hline
F &4.2 &36.0 &1027.0 &10 &-0.55 &3.0 &  -0.491  & 3.085  &-0.485  & 3.140 & -0.479 &  3.201  &-0.474&   3.250 \\ \hline
G &4.7 &36.0 &1027.0 &11 &-0.43 &3.0 & -0.458  & 3.397&  -0.452 &  3.458  &-0.446  & 3.526 & -0.441&  3.580 \\  \hline
H &5.4 &34.9 &1026.1 &13 &-0.35& 3.4&-0.408  & 3.767  &-0.402  & 3.835  &-0.397  & 3.910 & -0.392  & 3.971 \\ \hline
\end{tabular}
\caption{Theoretical results $T_i$ and $V$ and measurements for experiments (A-H) taken from Table 3 of ~\cite{KM2015}.
Values of other parameters used are
$\lambda=5.91 \times 10^{-2}$~K kg/g, $\beta=7.86\times 10^{-4}$~kg/g, $\rho_{s}=916.17$~kg/m$^3$, $c_w=4000$~J/(kgK), $c_s=2000$~J/(kgK), $L=3.34 \times 10^5$~J/kg, 
$\nu=1.805\times 10^{-6}$~m$^2$/s, $\kappa_S=7.034\times 10^{-10}$m$^2$/s, and $\kappa_T=1.357\times 10^{-7}$~m$^2$/s. }
\label{tab2}
\end{table*}
\end{widetext}

Kerr and McConnochie had developed a theoretical model to explain their experimental results~\cite{KM2015}. They expanded the idea of diffusive growth of boundary layers to get $\delta_T = \sqrt{\kappa_T \tau}$ and $\delta_S = \sqrt{\kappa_S \tau} + V \tau$ but they inconsistently estimated $\tau$ using  $\delta_S \approx \sqrt{\kappa_S \tau}$ and an empirical expression of $\Sh\equiv h/\delta_S$, $\Sh = \gamma \Ra_S^{1/3}$, where $\gamma$ is a constant, and obtained
\begin{equation}
\delta_S^{\rm KM} = \sqrt{\kappa_S \tau} \frac{S_a}{S_i} = \frac{1}{\gamma} \frac{h}{\Ra_S^{1/3}} \frac{S_a}{S_i} 
\label{deltaSKM}
\end{equation}
With Eq.~(\ref{deltaSKM}), their model in effect assumes
\begin{eqnarray}
R^{\rm KM} = \Le^{1/2} \frac{S_i}{S_a}, \qquad \Sh^{\rm KM} = \gamma^{\rm KM} \ \Ra_S^{1/3} \frac{S_i}{S_a}
 \label{KMtheory}
\end{eqnarray}
and employs Eq.~(\ref{KMtheory}) to estimate $S_i$ using Eq.~(\ref{eqSi2}) and $V$ up to a constant $\gamma^{\rm KM}$ using Eq.~(\ref{V}). Despite a good agreement of their model estimates of $V$ with the experimental measurements $V_m$ using a fitted value of $\gamma^{\rm KM} = 0.090 \pm 0.004$, their expression of $\Sh^{\rm KM}$ is not consistent with the general understanding that $\Sh$ should be a function of the control parameters of the flow, which depend on the characteristic scale $S_a-S_i$ and not $S_i$ or $S_a$ separately. Moreover, using $R^{\rm KM}$ in Eq.~(\ref{eqSi2}) results in a linear equation for $S_i$, which yields
\begin{equation}
S_i^{\rm KM} =  \frac{(L+c_s T_{\delta} - \Le^{1/2} c_w T_a)S_a}{\lambda c_w \Le^{1/2} S_a + L+ c_s T_{\delta}}
\label{SiKM}
\end{equation}
Thus, their model estimate of the salinity $S_i^{\rm KM}$ at the ice-water interface becomes negative and thus unphysical when $T_a$ exceeds $(L+c_s T_{\delta})/c_w\Le^{1/2}$. 

\subsection{Buoyancy-driven melt rate of an ice mass in seawater}

\begin{figure}[ht]
    \centering
   \includegraphics[width=1.05\linewidth]{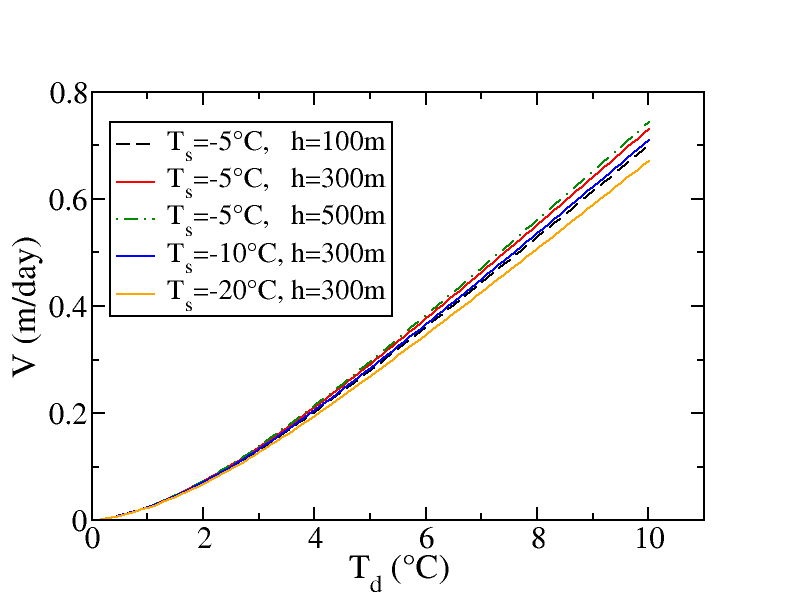}
    \caption{Theoretical estimates of buoyancy-driven melt rate $V$ of an ice mass with a near-vertical ice face of submerged height $h$ and interior temperature $T_s$ in seawater with ambient salinity $S_a = 34.0$~g/kg. }
    \label{Fig5}
\end{figure}

The theory that we have developed enables us to estimate the melt rate $V$ driven by buoyancy of an ice mass such as a tidewater glacier with a near-vertical ice face of submerged height $h$, interior temperature $T_s$ in seawater with ambient temperature $T_a$ and salinity $S_a$. We fix $S_a = 34.0$~g/kg, $\Sc=2.57 \times 10^3$, $\Pran=13.3$ and calculate $V$ as a function of the thermal driving measured by $T_d=T_a-T_L(S_a)$~\cite{JM1981} for different values of $T_s$ and $h$. The results are shown in Fig.~\ref{Fig5}. Our theoretical estimates of $V$ are significantly larger than the ambient melt estimates of 0.01 to 0.07~m/day for a tidewater glacier of $h \approx 200$~m in ocean with ambient temperatures of 1 to 4.5$^\circ$C, calculated using existing theory that employs melt parameterization with typical coefficient values~\cite{Sutherland2019}.

For $h \ge 100$~m, the values of $\Ra_S$ exceed $10^{19}$ and our theoretical estimates of $\Sh$ are well-described by $\Sh = \gamma \Ra_S^{1/3}$ with $\gamma \approx 0.021$. This $\Ra_S^{1/3}$-scaling of $\Sh$ with a  value of $\gamma$ much smaller than $\gamma^{\rm air}$ is in accord with the theoretical result for heat flux in turbulent vertical convection~\cite{Ching2023}. Substituting this $\Ra_S$-dependence of $\Sh$ in Eq.~(\ref{V}), we obtain
\begin{equation} 
V =  \frac{\gamma \rho_w}{\rho_s} \left( \frac{\beta g \kappa_S^2}{\nu} \right)^{1/3} \frac{(S_a-S_i)^{4/3}}{S_i} 
\label{V_Si}
\end{equation}
When $T_d=0$, Eq.~(\ref{eqSi}) implies that $S_i=S_a$ thus there is no salt and heat transfer from the ambient seawater to the ice face and $V=0$. For $T_d >0$, heat  transfers from the seawater to the vertical ice face causing it to melt. In the high-$\Ra_S$ limit, the flux ratio $R$ has a weak dependence on $\Ra_S$, then using Eq.~(\ref{eqSi}), it can be shown that $S_i$ decreases when $T_a$ is increased. Using Eq.~(\ref{V_Si}), it can be seen that $V$ increases when $S_i$ decreases. Hence, $V$ increases with $T_a$ or $T_d$.  For an ice mass with a higher $T_s$, less heat is transferred to the ice mass thus $V$ increases with $T_s$ when the other parameters are held fixed.  When $h$ increases, $\Ra_S$ generally increases and in the relevant range of $\Ra_S$, $R$ decreases slightly with $\Ra_S$. Using Eq.~(\ref{eqSi}), it can be shown that when $R$ decreases, $S_i$ decreases. Thus, $V$ increases with $h$ when the other parameters are held fixed. These results of $V$ increasing with $T_s$ or $h$ with the other parameters held fixed are confirmed and the increase of $V$ with $h$ is found to flatten for larger $h$~(see Fig.~\ref{Fig5}). Moreover, it can be seen in Fig.~\ref{Fig5} that $V$ is well approximated by a linear function of $T_d$ when $T_d > 3^\circ$C. 

Kerr and McConnochie had used their theoretical model to estimate $V$ for an ice mass of an interior temperature $T_s=-17^\circ$C in seawater of $S_a=34.0$~g/kg and showed that their estimated values can be fitted by $V = 7.8 T_d^{1.34}$ with $V$ in m/yr and $T_d$ in $^\circ$C~\cite{KM2015}.  As discussed above, their theoretical model suffers from an inconsistency in their estimate of $\delta_S$ or $\Sh$, leading to an expression of $\Sh^{\rm KM}$ that depends not only on the control parameter $\Ra_S$ but also separately on $S_i$. Moreover, the value of $\gamma^{\rm KM}$ (for $\Sc=2.56 \times 10^3$) is comparable to $\gamma^{\rm air}$, contrary to the expectation that the prefactor $\gamma$ should be decreasing with $\Sc$, which follows from the analogy of $\Sh$ to heat flux in turbulent vertical convection and the numerical finding that the latter decreases with $\Pran$ in a DNS study~\cite{HNVL2022}. Furthermore, their estimate of $S_i$ becomes negative and unphysical when $T_a$ exceeds $(L-c_sT_s)/c_w \Le^{1/2}=6.6^\circ$C.

\section{Conclusions}

Reliable estimates of the melt rate of tidewater glaciers require accurate knowledge of salt and heat fluxes at the near-vertical glacier face in contact with seawater.
In this paper, we have first developed a theory for salt and heat fluxes in turbulent double-diffusive vertical convection in an infinite channel driven mainly by the salinity difference, building upon a model for the space-dependent eddy salinity diffusivity, adopted from turbulent vertical convection~\cite{JFM}.  Our theory, which gives analytical estimates for the fluxes and their ratio $R$, explains the complex dependence of $R$ on the Lewis number $\Le$ observed in direct numerical simulation~\cite{HVL2023}. Specifically, $R$ depends not only on $\Le$ but also on the salinity Rayleigh number $\Ra_S$ and the Schmidt number $\Sc$, and for fixed values of $\Ra_S$ and $\Sc$, $R \sim \Le^{1/3}$ for small $\Le$, has a steeper dependence on $\Le$ for intermediate $\Le$ and eventually becomes independent of $\Le$ for very large $\Le$. Thus, the observed steeper dependence closer to $\Le^{1/2}$ for larger $\Le$~\cite{HVL2023} can be understood without the implied transition towards a $\Le^{1/2}$ scaling.

Then we have extended  our theory for the idealized system to attain theoretical estimate of buoyancy-driven
melt rate of a vertical ice surface in seawater. We have compared our theoretical estimates of the melt rate and the temperature (or salinity) of the ice surface with laboratory experiments of the melting of a vertical ice wall in homogeneous sodium chloride solution~\cite{KM2015}, and good agreement is found. For a realistic ice mass in seawater such as a tidewater glacier, factors other than buoyancy, e.g., ocean currents, wind-driven flows, internal gravity waves and surface wind waves, can drive the melting of the ice mass. Using our theory, we have estimated the buoyancy-driven melt rate of an ice mass with a near-vertical ice face in seawater with ambient salinity of $S_a \approx 34$~PSU and different ambient temperatures $T_a$. Our theoretical estimates of the buoyancy-driven melt rates are significantly larger than estimates obtained using an existing theory with melt parameterization with typical coefficient values~\cite{Sutherland2019}. Our results also show that the buoyancy-driven melt rate increases for glaciers with deeper submerged height and warmer interior temperature. It would be interesting to undertake careful and systematic comparison of the theoretical estimates with observations of tidewater glaciers in future studies.

\acknowledgments
The authors acknowledge support from the Hong Kong Research Grants Council (Grant No. CUHK 14302419).

\end{document}